\documentclass[12pt]{article} 
\def\cdate{{October 30, 2023}}
\usepackage{amsfonts,amsmath,amscd,amssymb,latexsym,mathptmx,txfonts} 
\usepackage{xcolor,graphics,framed}
\colorlet{shadecolor}{lightgray}
\usepackage[american]{babel}

\makeatletter
\renewcommand{\@evenhead}{\raisebox{0pt}[\headheight][0pt]{\vbox{\hbox
to \textwidth{\thepage\hfil\strut\textsc{\leftmark}}\hrule}}}
\renewcommand{\@oddhead}{\raisebox{0pt}[\headheight][0pt]{\vbox{\hbox
to \textwidth{\textsc{\rightmark}\hfil\strut\thepage}\hrule}}}
\makeatother
\makeatletter
\def\timenow{
\@tempcnta=\time \divide\@tempcnta by 60 \number\@tempcnta:\multiply
\@tempcnta by 60 \@tempcntb=\time \advance\@tempcntb by -\@tempcnta
\ifnum\@tempcntb <10 0\number\@tempcntb\else\number\@tempcntb\fi}
\newcounter{outputpage}
\renewcommand{\@oddhead}
{\stepcounter{outputpage}\hfill\hfill\theoutputpage}
\renewcommand{\@evenhead}
{\stepcounter{outputpage}\hfill\hfill\theoutputpage}
\renewcommand{\@oddfoot}
{\vbox{
\hrule
\vspace{3pt}
\hfil
{\scriptsize\textit{
\hfill\hfill\jobname.tex; \today; \timenow; p. \theoutputpage}}
\hfil
}}
\renewcommand{\@evenfoot}
{\vbox{
\hrule
\vspace{3pt}
\hfil
{\scriptsize\textit{
\hfill\hfill\jobname.tex; \today; \timenow; p. \theoutputpage
}}
\hfil
}}
\makeatother

\def\RR{{\mathbb R}}

\def\ZZ{{\mathbb Z}}

\def\cA{{\cal A}}
\def\cC{\mathcal{ C}}

\def\cM{{\cal M}}

\def\cV{{\cal V}}

\def\cP{{\cal P}}
\def\cR{\mathcal{ R}} 
\def\cA{{\cal A}}

\def\cC{{\cal C}}

\def\cM{{\cal M}}
\def\cV{{\cal V}}

\def\cP{{\cal P}}
\def\cR{{\cal R}}

\def\hfrak{{\mathfrak{h}}}

\def\tr{\mathrm{ tr\,}} 
\def\Tr{\mathrm{ Tr\,}}

\def\Det{\mathrm{ Det\,}}

\def\Real{\mathrm{Re\,}}

\def\dim{\mathrm{dim\,}}

\def\vol{\mathrm{ vol\,}}

\def\End{\mathrm{End}}

\def\nmt{{
\null
\vspace{-4cm}
\par
\hspace*{50truemm}{\hrulefill}
\par
\vskip-4truemm
\par
\hspace*{50truemm}{\hrulefill}
\par\vskip5mm
\par
\hspace*{50truemm}{{\large\sc 
New Mexico Tech {\rm 
(\cdate)
}}}\vskip4mm
\par
\hspace*{50truemm}{\hrulefill}
\par
\vskip-4truemm
\par
\hspace*{50truemm}{\hrulefill}
\par}}

\def\be{\begin{equation}} 
\def\ee{\end{equation}} 
\def\bea{\begin{eqnarray}} 
\def\eea{\end{eqnarray}} 
\def\bed{\begin{definition}{\ }}
\def\eed{\end{definition}}
\def\ed{\end{document}}
\def\bp{\begin{proposition}}
\def\ep{\end{proposition}}
\def\bc{\begin{center}}
\def\ec{\end{center}}
\def\bi{\begin{itemize}} 
\def\ei{\end{itemize}} 
\def\benum{\begin{enumerate}} 
\def\eenum{\end{enumerate}} 
\def\bmp{\begin{minipage}} 
\def\emp{\end{minipage}} 

\newtheorem{theorem}{Theorem}

\newtheorem{proposition}{Proposition}

\newtheorem{definition}{Definition}

\begin{document}

\begin{titlepage}
\thispagestyle{empty}
\nmt

\bigskip
\bigskip
\centerline{\huge\bf Spectral Asymptotics}
\bigskip
\centerline{\huge\bf of Elliptic Operators}
\bigskip
\centerline{\huge\bf on Manifolds}
\bigskip
\bigskip
\bigskip
\centerline{\Large\bf Ivan G. Avramidi}
\bigskip
\centerline{\it Department of Mathematics}
\centerline{\it New Mexico Institute of Mining and Technology}
\centerline{\it Socorro, NM 87801, USA}
\centerline{\it E-mail: ivan.avramidi@nmt.edu}
\bigskip
\medskip

\begin{abstract}

The study of spectral properties of natural geometric elliptic partial differential
operators  acting on smooth sections of vector
bundles over Riemannian manifolds
is a central theme in global analysis, differential geometry and
mathematical physics.  
Instead of studying the spectrum of a differential operator $L$ directly one usually
studies its spectral functions, that is, spectral traces of some functions of
the operator, such as the spectral zeta function 
$\zeta(s)=\Tr L^{-s}$ 
and the heat trace $\Theta(t)=\Tr\exp(-tL)$.
The kernel $U(t;x,x')$ of the heat semigroup $\exp(-tL)$, called the heat kernel, plays
a major role in quantum field theory and quantum gravity, 
index theorems, non-commutative geometry,
integrable systems and financial mathematics.
We review some recent progress in the study of spectral asymptotics. 
We study more general spectral functions, such as $\Tr f(tL)$, that we call
quantum heat traces.
Also, we define
new invariants 
of differential operators that depend not only on the their eigenvalues but also on the
eigenfunctions, and, therefore, contain much more information about the geometry
of the manifold. Furthermore, we study some new invariants, such as
$\Tr\exp(-tL_+)\exp(-sL_-)$, that contain relative
spectral information of two differential operators.
Finally we show how the convolution of the semigroups of two different operators
can be computed by using purely algebraic methods.

\end{abstract}

\end{titlepage}


\section{Introduction}
\setcounter{equation}0 


The study of spectral properties of natural geometric partial differential
operators is a central theme in global analysis, differential geometry and
mathematical physics. In particular, the basic question of spectral geometry is:
``{To what extent does the spectrum of an elliptic partial differential operator
determine the geometry of the underlying manifold?}'', or as M. Kac put it in
his famous paper \cite{kac66}: ``{Can one hear the shape of a drum?}'' In
general, the answer to Kac's question is ``no'' \cite{sunada85,schueth99}.
Instead of studying the spectrum of a differential operator directly one usually
studies its spectral functions, that is, spectral traces of some functions of
the operator, such as the spectral zeta function and the heat trace. The heat
trace is the trace of the heat kernel, which is the fundamental solution of the
heat equation for an elliptic partial differential operator with a positive
leading symbol (see, e.g. \cite{gilkey95, vassilevich03, berline92, hurt83, kirsten01, 
grigoryan12, chavel84, avramidi00, avramidi10, avramidi15}). 
The existence of non-isometric isospectral manifolds
demonstrates that the spectrum alone does not determine the geometry. That is
why, it makes sense to study more general invariants of partial differential
operators, maybe even such invariants that are not spectral invariants, that is,
invariants that depend not only on the eigenvalues but also on the
eigenfunctions, and, therefore, contain much more information about the geometry
of the manifold \cite{avramidi14a,avramidi17,avramidi20a,avramidi20b}.

Another motivation to study the heat kernel comes from quantum field theory and
statistical physics. The main objects of interest are the effective action (or
the partition function) and the Green functions (or the correlation functions)
\cite{landau69,dewitt75,birrell80,wald94,parker09}.
All these objects are expressed in terms of the functional determinants of some
self-adjoint elliptic partial differential operators, their heat traces and
their resolvents. It turns out that all of them can be expressed in terms of the
heat kernels of those operators.


In financial mathematics one uses stochastic differential equations to model the
random behavior of some financial assets. Then the behavior of the corresponding
derivative securities is determined by deterministic parabolic partial
differential equation (such as the diffusion or heat equation) with an elliptic
partial differential operator of second order. The conditional probability
density is then nothing else but the heat kernel (for a brief introduction
see \cite{avramidi15} with more references therein).


Many problems in mathematics and physics naturally lead to the presence of
boundaries and to the corresponding boundary value problems for partial
differential operators.  The type of the boundary conditions is not limited to
the classical Dirichlet and Neumann ones. In some applications, such as quantum
gravity and applied mathematics, there appear mixed boundary conditions on
vector bundles (that mix the Dirichlet and the Neumann one), oblique boundary
conditions (that involve the tangential derivatives to the boundary) 
\cite{avramidi98c,avramidi99a,avramidi06}, and even
discontinuous boundary conditions, so called Zaremba boundary conditions, 
(that jump from Dirichlet to Neumann across a
co-dimension $2$ submanifold in the boundary) \cite{avramidi04a,seeley02}.


Most natural elliptic partial differential operators 
are second order operators with scalar leading symbols,
so called Laplace type operators, or first order operators
whose square is a Laplace type operator, so called Dirac type
operators.  
However, in some applications, such as in gauge field theories and quantum gravity, 
there appear second-order
elliptic partial differential operators with non-scalar leading symbols, 
so called non-Laplace type operators
\cite{avramidi01,avramidi06}. Another motivation for studying 
such operators is non-commutative
geometry and matrix general relativity
where the metric tensor, that is, the inner product in the tangent
bundle, is endomorphism valued in some vector bundle
\cite{avramidi04b,avramidi04c,avramidi05,avramidi06}.


In the generic situation when it is impossible to compute the heat kernel
exactly, it becomes very important to study highly symmetric spaces
such as Lie groups and symmetric spaces
\cite{camporesi90,hurt83,avramidi93b,avramidi21,avramidi09,avramidi23}
or
various asymptotic regimes. 
Of special interest
is the study of the short-time asymptotic expansion of the heat kernel
\cite{gilkey75,gilkey95,berline92, chavel84, avramidi91,avramidi10}. This
expansion is closely related to the semi-classical expansion in quantum theory
and the high-temperature expansion in statistical physics
\cite{dewitt75,birrell80}. The coefficients of
this expansion, called the heat invariants, are spectral invariants associated
with the asymptotic properties of the spectrum. There are also non-trivial links
between spectral invariants and non-linear completely integrable systems, such
as the Korteweg-de Vries hierarchy
\cite{perelomov76,olshanetsky81,olmedilla81,dickey03,hurt83,iliev05,avramidi00mn,avramidi14jmp}. 
In many interesting cases such systems are,
in fact, infinite-dimensional Hamiltonian systems, and the infinite set of
integrals of motion of these systems is related to the spectral invariants of a
linear elliptic partial differential operator
(see \cite{dickey03,avramidi14jmp}).


\section{Heat Kernel} 
\setcounter{equation}0 


\subsection{Elliptic Operators}

Let $(M,g)$ be a smooth compact Riemannian manifold 
of dimension $n$ equipped with a positive definite Riemannian metric $g$. We denote
the local coordinates by $x^\mu$, with Greek indices running over $1,...,n$. The
Riemannian volume element is defined as usual by $d\vol = dx\;g^{1/2}$, where $g
= \det g_{\mu\nu}$ and $dx=dx^1\wedge \dots\wedge dx^n$ is the standard Lebesgue
measure.

Let ${\cal V}$ be a smooth vector bundle over $M$ with the typical fiber $V$ of
dimension $N$. Let $C^\infty(\cV)$ be the space of smooth sections of the bundle
$\cV$. The completion of the space
$C^\infty(\cV)$ defines the Hilbert space $L^2(\cV)$ of square integrable
sections.
Let 
\be \nabla^\cV: C^\infty(\cV)\to C^\infty(T^*M\otimes \cV) 
\ee
 be a
compatible connection on the vector bundle ${\cal V}$. By using the Levi-Civita
connection of the metric $g$ the connection is given its unique natural
extension to bundles in the tensor algebra over $\cV$, its dual $\cV^*$ and the
tangent and cotangent bundles, $TM$ and $T^*M$; the resulting connection will
usually be denoted just by $\nabla$.
The fiber inner product $\langle\;,\;\rangle$ defines a natural $L^2$ inner
product, $(\;,\;)$, on the bundle $\cV$ and the $L^2$-trace, $\Tr$, using the
invariant Riemannian measure on the manifold $M$. 


Let 
\be
\nabla^*:  C^\infty(TM\otimes \cV) \to C^\infty(\cV)
\ee
be the formal adjoint of the connection.
Let $a$, $B$ and $Q$ be smooth maps
\bea
&& a: T^*M\otimes \cV\to TM\otimes\cV,
\\
&& B: T^*M\otimes \cV\to \cV,
\\
&& Q: \cV\to \cV.
\eea
defined by endomorphism-valued tensors satisfying
\be
a^{\mu\nu}=a^{\nu\mu},
\ee
\be
(a^{\mu\nu})^*=a^{\mu\nu},
\qquad
(B^\mu)^*=-B^\mu,
\qquad
Q^*=Q
\ee
Every formally 
self-adjoint second-order partial differential operator 
\be
L: C^\infty(\cV)\to C^\infty(\cV)
\ee
has the form
\bea
L &=& 
-\nabla_\mu a^{\mu\nu}\nabla_\nu + B^\mu\nabla_\mu + \nabla_\mu B^\mu + Q.
\label{oper}
\eea

Natural non-Laplace type operators can be constructed as follows.
Let 
\be
\Gamma:  T^*M\otimes \cV\to \cV
\ee 
be a smooth map; this defines the
Dirac type operator
\be
D=\Gamma\nabla : C^\infty(\cV)\to C^\infty(\cV)
\ee
and the operator
\be
L=D^*D: C^\infty(\cV)\to C^\infty(\cV);
\ee
in this case
\be
a^{\mu\nu}=\frac{1}{2}\left(\Gamma^{\mu\; *} \Gamma^\nu
+\Gamma^{\nu\; *} \Gamma^\mu
\right).
\ee

More generally,
let $\cV_j$, $j=1,\dots,s$, be some  vector bundles and
\be
P_j: T^*M\otimes \cV\to \cV_j
\ee
be some smooth maps. This defines the gradients 
\be
G_j=P_j\nabla: C^\infty(\cV)\to C^\infty(\cV_j),
\ee
and the operator
\be
L=\sum_{j=1}^s \alpha_j G_j^*G_j: C^\infty(\cV)\to C^\infty(\cV),
\ee
where $\alpha_j$ are some real constants.
In this case
\be
a^{\mu\nu}=\sum_{j=1}^s \alpha_j 
\frac{1}{2}\left(P_j^{\mu\; *} P_j^\nu
+P_j^{\nu\; *} P_j^\mu
\right).
\ee

The leading symbol of the operator $L$ is given by
the endomorphism
\be
H(x,\xi)=a^{\mu\nu}(x)\xi_\mu\xi_\nu,
\ee
with $x\in M$ and $\xi\in T_x^*M$.
Since this matrix is self-adjoint all its eigenvalues must be real.
The operator $L$ is elliptic if all eigenvalues are positive, in other words,
the matrix $H(x,\xi)$ is positive. 
In the case when the manifold $M$ is closed, that is, compact without boundary,
the operator $L$ is also self-adjoint;
strictly speaking, it is essentially self-adjoint, that is, it has a unique
self-adjoint extension (from now on, we will just say that the operator $L$ is
self-adjoint).

If all eigenvalues of the leading symbol of the operator $L$ are equal, that is,
the operator $L$ has a scalar positive definite leading symbol defined by the
Riemannian metric
\be
H(x,\xi)=|\xi|^2 I,
\ee
with $|\xi|^2=g^{\mu\nu}(x)\xi_\mu\xi_\nu$,
then the operator is called of Laplace type. In this case
the connection can be redefined to absorb the vector $B^\mu$, so that
 every Laplace type operator has the form
\be
L=-\Delta+Q,
\ee
where
\bea
\Delta &=& 
g^{\mu\nu}\nabla_\mu\nabla_{\nu}
\nonumber\\
&=& g^{-1/2}(\partial_\mu+{\cal A}_\mu)g^{1/2}g^{\mu\nu}
(\partial_\nu+{\cal A}_\nu)
\eea
is the Laplacian.
Here and below $\partial_\mu=\partial/\partial x^\mu$ denotes the partial derivative.
A Laplace type operator is defined in terms of three pieces of local 
information: the Riemannian metric $g$, the connection one-form ${\cal A}$ 
and the potential $Q$.


\subsection{Heat Kernel}

Let $L$ be a self-adjoint elliptic positive partial differential operator of
second order on the Hilbert space $L^2(\cV)$. For manifolds with boundary the
domain of the operator $L$ has to be supplemented with some suitable elliptic
boundary conditions. For compact manifolds the spectrum of the operator $L$ is
an increasing real sequence of eigenvalues $\{\lambda_k\}_{k\in\ZZ_+}$ with the
corresponding orthonormal eigensections $\{\varphi_k\}_{k\in\ZZ_+}$ (counted
with multiplicities) defined by
\cite{gilkey95,berline92}
\be
(L-\lambda_k)\varphi_k=0,
\qquad
(\varphi_i,\varphi_j)_{L^2}=\delta_{ij}.
\ee 
For noncompact manifolds
the spectrum of the operator $L$ is continuous,
it goes from a positive real constant $c$ to $\infty$.
In general, it is impossible to compute the spectrum exactly.
That is why, it becomes of special importance the
study of the asymptotics of the eigenvalues (and the eigensections)
as $k\to \infty$. Rather than doing this directly it is more convenient
to study the asymptotics of some spectral functions and special traces
such as the heat trace and the zeta function
\cite{gilkey95,berline92,kirsten01}.

Let $\cV\boxtimes\cV^*$
be the external tensor product of the bundles $\cV$ and $\cV^*$
over the product manifold $M\times M$.
The heat kernel $U(t;x,x')$ of the operator $L$ is a one-parameter family
of smooth sections of $\cV\boxtimes\cV^*$
defined by requiring it to
satisfy the heat equation
\be
(\partial_t+L)U(t;x,x')=0
\label{14mus}
\ee
for $t>0$ 
with the initial condition
\be
U(0^+;x,x')=\delta(x,x')\,,
\ee
that is,
\be
U(t;x,x')=\exp(-tL)\delta(x,x')\,,
\ee
where $\delta(x,x')$ is the covariant Dirac delta-distribution.

The heat kernel is regular
at the diagonal with a well defined
diagonal value \cite{gilkey95}
\be
U^{\rm diag}(t;x)=U(t;x,x),
\ee
which is a section of the endomorphism
bundle $\End(\cV)$.
The heat kernel diagonal is well defined on both compact and noncompact
manifolds.
For compact manifolds
the heat kernel can be computed in terms of the spectral data
\bea
U(t;x,x') &=& 
\sum_{k=1}^\infty e^{-t\lambda_k}\varphi_k(x)\varphi^*_k(x').
\label{229igaz}
\eea
and has a well defined heat trace $\Theta(t)=\Tr\exp(-tL)$,
\bea
\Theta(t) &=&\sum_{k=1}^\infty e^{-t\lambda_k}
\nonumber\\
&=&\int\limits_M d\vol\; \tr U^{\rm diag}(t);
\eea
here and below 
$\tr$ denotes the fiber trace.
This also enables one to define the spectral zeta function
by the Mellin-Laplace transform of the heat trace
\be
\zeta(s,\lambda)
=\sum_{k=1}^\infty \frac{1}{(\lambda_k-\lambda)^s}
=\frac{1}{\Gamma(s)}\int\limits_0^\infty t^{s-1}
e^{t\lambda}\Theta(t),
\ee
where $\lambda$ is a complex parameter with a
sufficiently large negative real part,
and the regularized spectral determinant
\cite{gilkey95,kirsten01}
\be
\Det(L-\lambda)=\exp\left\{-\partial_s\zeta(s,\lambda)\big|_{s=0}\right\}.
\ee


The resolvent $G(\lambda;x,x')$ of the operator $L$ is a section of  
$\cV\boxtimes\cV^*$ depending on a complex parameter
$\lambda$ defined by
\be
(L-\lambda)G(\lambda;x,x')=\delta(x,x').
\label{12mus}
\ee
The resolvent is related to the heat kernel by the
Laplace transform
\bea
G(\lambda;x,x') &=& \int\limits_{0}^{\infty}
dt\; e^{t\lambda}U(t;x,x').
\label{224igaz}
\eea 
For compact manifolds there is the spectral representation
of the resolvent
\bea
G(\lambda;x,x') &=& 
\sum_{k=1}^\infty \frac{1}{\lambda_k-\lambda}\varphi_k(x)\varphi^*_k(x').
\label{229igazm}
\eea 
Off diagonal, that is, for $x\ne x'$,
the resolvent is an analytic function of $\lambda$ for 
$\Real\lambda<c$ with sufficiently
large negative real constant $c$.
It has a diagonal singularity as $x\to x'$.


On a general manifold the heat kernel cannot
be computed exactly, and that is why the short-time asymptotic
expansion as $t\to 0$ of the heat kernel and its heat trace
is studied.
For Laplace type operators it has a rather simple form.
Let $x'$ be a fixed point in the interior of the manifold;
we consider a
sufficiently small geodesic ball centered at $x'$, so
that each
point $x$ of the ball
can be connected by a unique geodesic
with the point $x'$. This can be always
done if the size of the ball is smaller than the injectivity radius of the
manifold
$r_{\rm inj}(M)$.

Let $\sigma=\sigma(x,x')$  be the Ruse-Synge function defined by
\cite{ruse31,synge60}
\be
\sigma(x,x')=\frac{1}{2}r^2(x,x'),
\ee
where
$r(x,x')$ is the geodesic distance between the points $x$ and $x'$,
and $\Delta(x,x')$
be the Van Vleck-Morette determinant
\be
\Delta(x,x')=g^{-1/2}(x)
\det\left\{-\frac{\partial^2\sigma(x,x')}{\partial x^\mu\partial x^{\nu'}}\right\}g^{-1/2}(x').
\ee
Near the diagonal of $M\times M$ these two-point functions are smooth
single-valued
functions of the coordinates of the points $x$ and $x'$
\cite{synge60,avramidi91,avramidi00,avramidi10}.

In the interior of the manifold (on a finite distance from the boundary, if present)
there is a local short time asymptotic expansion of the heat kernel 
of a Laplace type operator as $t\to 0^+$
\cite{gilkey95,dewitt75,avramidi00}
\be
U(t;x,x')\sim (4\pi t)^{-n/2}\Delta^{1/2}(x,x')
\exp\left\{-\frac{\sigma(x,x')}{2t}\right\}
\sum_{k=0}^{\infty} t^{k} a_{2k}(x,x'),
\label{hkoffdiag}
\ee
where $a_{k}(x,x')$ are the so-called local off-diagonal heat kernel coefficients.
Notice that there are no odd-order coefficients here, that is,
$a_{2k+1}(x,x')=0$.
The heat kernel coefficients can be computed in form of a 
covariant Taylor series from the recurence relations obtained 
by substituting this ansatz in the heat equation 
\cite{avramidi91}
with the initial condition
\be
a_0(x,x')=\cP(x,x'),
\ee
where $\cP(x,x')$ is the operator of parallel transport of sections 
along the geodesic from the point $x'$ to the point $x$
(for details, see \cite{avramidi91,avramidi00}). 

This expansion immediately gives the asymptotic expansion as $t\to 0^+$ 
of the heat kernel diagonal
\be
U^{\rm diag}(t;x) \sim (4\pi t)^{-n/2}\sum_{k=0}^{\infty}
t^{k}a^{\rm diag}_{2k}(x,x),
\ee
where $a_k^{\rm diag}(x)=a_k(x,x)$  are the diagonal local heat kernel coefficients.

For manifolds without boundary this also gives
the asymptotic expansion of the heat trace
\cite{greiner71,seeley69,gilkey95}
\be
\Theta(t)
\sim
(4\pi t)^{-n/2}\sum_{k=0}^\infty t^{k} A_{2k},
\label{239fa}
\ee
where 
\be
A_{2k}=\int\limits_M d\vol\; \tr a^{\rm diag}_{2k}
\label{240fa}
\ee
are the global heat kernel coefficients.
The heat trace, as well as the global heat kernel coefficients,
are obviously spectral invariants
of the operator $L$.
These coefficients were computed up to $a_8^{\rm diag}$ (see \cite{gilkey75,avramidi91,avramidi00}).
The low-order asymptotics of the heat trace of a Laplace type operator has the form
\bea
\Theta(t)
&=& (4\pi t)^{-n/2}
\Biggl\{N\vol(M)
+t\int\limits_M d\vol\left(
\frac{N}{6}R-\tr Q\right)
+O\left(t^{2}\right)\Biggr\},
\eea
where $R$ is the scalar curvature.


\section{Boundary Value Problems}
\setcounter{equation}0 

\subsection{Mixed Boundary Value Problem}

Now, let $M$ be a compact Riemannian manifold with smooth
boundary $\partial M$. 
We denote
the local coordinates on the boundary $\partial M$ by $x^i$, with Latin indices running over $1,...,(n-1)$. The
Riemannian volume element is defined as usual by $d\vol = d\hat x\;\hat g^{1/2}$, where 
$d\hat x=d\hat x^1\wedge \dots\wedge d\hat x^{n-1}$ is the standard Lebesgue
measure and $\hat g = \det \hat g_{ij}$, with $\hat g_{ij}$ the induced metric on the boundary.

Let  
$N$ be the inward-pointing unit normal vector field to the boundary $\partial
M$. To make the operator $L$ elliptic we need to impose some
boundary conditions on the boundary data.
The classical boundary conditions are
\bea
\varphi\big|_{\partial M} &=& 0, \qquad \mbox{(Dirichlet)},
\\
\nabla_N\varphi\big|_{\partial M} &=& 0, \qquad \mbox{(Neumann)}.
\eea
The Robin boundary condition is a slight generalization of the Neumann
one
\be
\left(\nabla_N+S\right)\varphi\Big|_{\partial M}=0,
\qquad \mbox{(Robin)}.
\ee
where $S$ is a smooth self-adjoint endomorphism.
One can go further and mix these boundary conditions as follows
\bea
(I-\Pi)\varphi\big|_{\partial M} &=& 0\,,\qquad
\\
\Pi\left(\nabla_N+S\right)\Pi\varphi\big|_{\partial M} &=& 0,
\eea
where $\Pi$ is a self-adjoint projection.
A Laplace type operator $L$ equipped with mixed boundary conditions is essentially
self-adjoint and elliptic
\cite{gilkey95,branson99,avramidi99a,avramidi04a}.

The heat kernel asymptotics as $t\to 0$ does not depend
on the boundary conditions
in the interior of the manifold and has the same form 
as the heat kernel for manifolds without boundary.
Near the boundary there is a narrow strip of the width of order $t^{1/2}$ where
one should add an additional term, whose role is to satisfy the boundary conditions.
In a narrow strip near the boundary
this compensating term in the heat kernel diagonal 
behaves as a distribution near the boundary as $t\to 0$
\cite{avramidi04a}. 
It is precisely this
feature that leads to the presence of boundary terms in the global heat kernel coefficients.
The heat trace
asymptotic expansion as $t\to 0^+$ has the half-integer
powers of $t$, 
\cite{greiner71,seeley69,branson90,gilkey95,vassilevich03,avramidi04a}
\be
\Theta(t)
\sim(4\pi t)^{-n/2}\sum_{k=0}^\infty t^{k/2}A_{k};
\label{8}
\ee
here
\be
A_{k}=\int\limits_M d\vol\; \tr\, a_{k}^{\rm diag}
+\int\limits_{\partial M} d\vol b_{k}\,,
\ee
where the boundary heat kernel coefficients $b_k$ are local invariants constructed 
polynomially from the jets of the symbols of both the operator $L$ and the boundary operator.

The low-order asymptotics have the form
\cite{branson90,branson99,avramidi04a}
\bea
\Theta(t)
&=& (4\pi t)^{-n/2}
\Biggl\{N\vol(M)
+t^{1/2}\, \int\limits_{\partial M}d\vol\frac{\sqrt{\pi}}{2}
\left(2\tr\Pi-N\right)
\\
&&
+t\Biggl[\int\limits_M d\vol\left(
\frac{N}{6}R-\tr Q\right)
+\int\limits_{\partial M}d\vol
\left(\frac{N}{3}K+2\tr\Pi S\right)
\Biggr]
+O\left(t^{3/2}\right)\Biggr\},
\nonumber
\eea
where $K$ is the trace of the extrinsic curvature of the boundary.

\subsection{Grubb-Gilkey-Smith Boundary Value Problem}

The mixed boundary conditions described above contain only normal derivative.
More general boundary conditions contain also tangential derivatives
along the boundary
\cite{gilkey83,seeley02,avramidi98c,avramidi99a}. 
Let $\hat\nabla_i$ be the tangential covariant derivative,
$\Gamma^i$ be a vector-valued anti-self-adjoint endomorphism
 and 
$\Lambda$ be a first-order formally self-adjoint tangential
differential operator defined by
\be
\Lambda=\frac{1}{2}\left(\Gamma^i\hat\nabla_i+\hat\nabla_i\Gamma^i\right)+S\,.
\ee
The Grubb-Gilkey-Smith boundary conditions 
(also called oblique boundary conditions)
then read
\cite{gilkey83}
\bea
(I-\Pi)\varphi\Big|_{\partial M} &=&0\,,\qquad
\\
\Pi\left(\nabla_N+\Lambda\right)\Pi\varphi\Big|_{\partial M} &=&0.
\label{3} 
\eea

A Laplace type operator $L$ equipped with such boundary conditions is essentially
self-adjoint but {\it not necessarily elliptic}. 
To be elliptic the operator $\Lambda$
has to satisfy the strong ellipticity condition
\cite{gilkey83,booss93,grubb96,gilkey95}. Let $T$ be a matrix
defined by the leading symbol of the operator $\Lambda$,
\be
T(\hat\xi)=\Gamma^j\hat\xi_j,
\ee
where $\hat\xi\in T^*\partial M$ is a covector on the boundary.
Since the matrices $\Gamma^i$ are anti-self-adjoint, the matrix
$T^2(\xi)$ is self-adjoint and negative, $T^2(\hat\xi)<0$, for any $\hat\xi\ne 0$.
Then the oblique boundary value problem is elliptic if the matrix
\be
I|\hat\xi|^2+T^2(\hat\xi)>0
\ee
is positive for any $\hat\xi\ne 0$.
Here $|\hat\xi|^2=\hat g^{ij}\hat\xi_i\hat\xi_j$ is defined with the metric
$\hat g^{ij}$ 
on the boundary.

When the boundary value problem is elliptic the heat trace 
asymptotic expansion
has the canonical form
(\ref{8}).
Contrary to the classical boundary value problems (Dirichlet, Neumann, mixed),
because of the non-commutativity of the matrices $\Gamma^i$, the explicit
form of the coefficients $b_k$ is unknown, in general;
the low-order asymptotic expansion has the form 
\cite{avramidi99a,avramidi98c}
\be
\Theta(t)
=(4\pi t)^{-n/2}\left\{N\vol(M)
+t^{1/2}\, \int\limits_{\partial M}d\vol
{\sqrt\pi\over 2}\Bigl(2\tr\Pi
-3N
+2\gamma
\Bigr)
+O(t)\right\},
\ee
where
\be
\gamma=\int\limits_{{\RR}^{n-1}} {d\hat\xi \,\over \pi^{(n-1)/2}}\,
\tr\exp\left(-|\hat\xi|^2-T^2(\xi)\right).
\label{3.45ee}
\ee
This can be computed explicitly in special cases.
If the matrices $\Gamma^i$ commute then 
\be
\gamma=\tr \left(I+\Gamma^2\right)^{-1/2},
\ee
where $\Gamma^2=\hat g_{ij}\Gamma^i\Gamma^j$. One can also
compute the coefficient $\gamma$ explicitly in the non-commutative case
when the matrices $\Gamma^i$ form a Clifford algebra
\be
\Gamma^i\Gamma^j+\Gamma^j\Gamma^i=-2\varkappa\Pi \hat g^{ij}
\ee
with a real parameter $\varkappa$.
This problem is elliptic if $\varkappa<1$ and
we obtain
\be
\gamma=(1-\varkappa)^{-(n-1)/2}\tr \Pi.
\ee


\subsection{Zaremba Boundary Value Problem}


Let the boundary $\partial M$ be decomposed as a
disjoint union $\partial M=\Sigma_1\cup \Sigma_2\cup\Sigma_0$, where 
$\Sigma_1$ and  $\Sigma_2$ are smooth compact submanifolds of $\partial M$ of
dimension $(n-1)$, with the boundary
$\Sigma_0$ which is a smooth compact
manifold without boundary of dimension $(n-2)$. 
Zaremba boundary value problem is defined by the following 
boundary conditions
\cite{avramidi04a,seeley02}
\bea
\varphi\big|_{\Sigma_1} &=& 0,\qquad
\\ 
(\nabla_N+S)\varphi\big|_{\Sigma_2} &=& 0\,,
\eea
where
$S$ is a smooth self-adjoint endomorphism.

Since the boundary operator is discontinuous, this problem is a singular
boundary value problem.
It is well known that 
in this case the heat trace asymptotic expansion as $t\to 0^+$ contains, in general, 
logarithmic terms
\cite{grubb96,gil03}
\be
\Theta(t)
\sim (4\pi t)^{-n/2}\sum_{k=0}^\infty t^{k/2}A_k
+\log t\sum_{k=0}^\infty t^{k/2}H_k\,.
\ee
However, Seeley \cite{seeley02} has shown that for Zaremba problem the logarithmic terms
do not appear, that is, all 
\be
H_k=0.
\ee 
Therefore, the heat trace asymptotics has the canonical form
(\ref{8}). However, the global heat kernel coefficients get a
contribution from the codimension 2 submanifold $\Sigma_0$
\cite{avramidi04a,seeley02}
\be
A_k=\int\limits_M d\vol \tr\, a_k^{\rm diag}
+\int\limits_{\Sigma_1} d\vol b_k^{(1)}
+\int\limits_{\Sigma_2} d\vol b_k^{(2)}
+\int\limits_{\Sigma_0} d\vol c_k\,.
\ee

It turns out \cite{avramidi04a,seeley02} that 
the
boundary conditions on the open  sets $\Sigma_1$ and $\Sigma_2$ are not
enough to fix the problem,  and an additional boundary condition along
the singular set $\Sigma_0$ is needed. This additional boundary condition
can be considered formally  as an extension of Dirichlet conditions from
$\Sigma_1$ to $\Sigma_0$,  (regular boundary condition)
\be
\left(\sqrt{\rho}\varphi\right)\Big|_{\Sigma_0}=0,
\label{regbc-r}
\ee
where  $\rho$ is the normal geodesic distance to the singular set $\Sigma_0$,
or an extension of Neumann (or Robin) conditions from
$\Sigma_2$ to $\Sigma_0$,
\be
(\partial_\rho-h)\left(\sqrt{\rho}\varphi\right)\Big|_{\Sigma_0}=0\,,
\label{singbc-r}
\ee
where $h$ is a real parameter.
However, strictly speaking the boundary
condition on $\Sigma_0$  does not follow from the boundary conditions on
$\Sigma_1$ and $\Sigma_2$  and can be chosen rather arbitrarily. 

The coefficients of the asymptotic expansion 
can be computed by constructing asymptotic solutions
\cite{avramidi04a}: 
\begin{enumerate}
\item
in the interior,
\item
in a thin shell near the $\Sigma_1$ and $\Sigma_2$, and 
\item
in a thin strip close to the singular submanifold $\Sigma_0$. 
\end{enumerate}
The trace of the heat kernel of the Zaremba boundary value problem
has the following asymptotic expansion as $t\to 0^+$
\bea
\Theta(t)
&=&(4\pi t)^{-n/2}
\Biggl\{
N\vol(M)
+t^{1/2}{\sqrt{\pi}\over 2}N 
\left[\vol(\Sigma_2)-\vol(\Sigma_1)\right]
\nonumber\\
&&
+t\Biggl[\int\limits_M d\vol \left({N\over 6}R-\tr Q\right)
+{N\over 3} \int\limits_{\Sigma_1} d\vol K
+\int\limits_{\Sigma_2} d\vol \left({N\over 3}  K
+2\tr S
\right)
\Biggr]
\nonumber\\
&&
+\alpha\, {\pi\over 4} N \vol(\Sigma_0) 
+O\left(t^{3/2}\right)
\Biggr\}\,,
\eea
where $\alpha=-1$ for the boundary condition (\ref{regbc-r}) and
$\alpha=7$ for the boundary condition (\ref{singbc-r})
(for the details of the calculation see \cite{avramidi04a}).


\section{Non-Laplace Type Operators}
\setcounter{equation}0 


We consider a non-Laplace type operator of the form
\be
L=-\nabla_\mu a^{\mu\nu}\nabla_\nu+Q,
\label{genlform}
\ee
where $a^{\mu\nu}$ is an endomorphism-valued symmetric tensor.
If we assume, in addition, that the leading symbol of the operator $L$ is positive
then the operator $L$ is self-adjoint and elliptic
and, therefore, has the same canonical heat trace expansion
of the form
(\ref{8})
\begin{equation}
\Theta(t)
\sim (4\pi t)^{-n/2}\sum_{k=0}^\infty 
t^{k/2}A_{k}.
\end{equation}


For manifolds without boundary  one can get
the asymptotic expansion of the heat kernel as follows
\cite{avramidi01}.
Let $\xi\in T^*M$ be a covector
and  $\langle \xi,x\rangle =\xi_\mu x^\mu$.
It is easy to see that
\be
\exp\left(-i\langle \xi,x\rangle\right)L\exp\left(i\langle \xi,x\rangle\right)
=H+K+L,
\ee
where
\be
H(x,\xi)=a^{\mu\nu}(x)\xi_\mu\xi_\nu
\ee
is the leading symbol of the operator $L$
and $K$ is the first order operator defined by
\be
K=-i\xi_\mu \left(a^{\mu\nu}\nabla_\nu
+\nabla_\nu a^{\mu\nu}\right).
\ee
Then the asymptotics of the heat trace as $t\to 0$ are given by
\begin{equation}
\Theta(t)
\sim (4\pi t)^{-n/2}\int\limits_M dx\int\limits_{\RR^n} {d\xi\over \pi^{n/2}} 
\tr\exp\left(-H-\sqrt tK-tL\right)\cdot I\,,
\end{equation}
By using the
Volterra series for the heat semigroup
we get 
\begin{eqnarray}
&&\exp\left(-H-\sqrt tK-tL\right)
=e^{-H}-
t^{1/2}\int\limits_0^1 d\tau_1 e^{-(1-\tau_1)H}K e^{-\tau_1 H}
\nonumber\\
&&
\qquad
+\,t\Biggl[ \int\limits_0^1d\tau_2\int\limits_0^{\tau_2}d\tau_1
  e^{-(1-\tau_2)H} K e^{-(\tau_2-\tau_1)H}Ke^{-\tau_1 H}-
\nonumber\\
&&
\qquad
\hphantom{+\,t\Biggl[}
  -\int\limits_0^1 d\tau_1 e^{-(1-\tau_1)H}
L e^{-\tau_1 H} \Biggr]+
O(t^2)\,.
\end{eqnarray}
Since $K$ is linear in $\xi$, the term proportional to $t^{1/2}$
vanishes after integration over $\xi$ and we obtain the first two
coefficients of the asymptotic expansion of the
heat kernel diagonal
in the form
\cite{avramidi01}
\bea
A_0 &=&\int\limits_M dx\int\limits_{\RR^{n}}\frac{d\xi}{\pi^{n/2}}\tr
\exp\left[-H(x,\xi)\right],
\\[5pt]
A_1 &=& 0,
\\
A_2&=&\int\limits_M dx\,\int\limits_{\RR^n}{d\xi\over \pi^{n/2}}\,
\tr\,\Biggl[
\int\limits_0^1d\tau_2\int\limits_0^{\tau_2}d\tau_1 e^{-(1-\tau_2)H}
K e^{-(\tau_2-\tau_1)H}Ke^{-\tau_1 H}-
\nonumber\\&&
     \hphantom{\int\limits_M dx\,\int\limits_{\RR^n}
{d\xi\over \pi^{n/2}}\,\tr_V\,\Biggl[}
-\int\limits_0^1 d\tau_1 e^{-(1-\tau_1)H}
L e^{-\tau_1 H} \Biggr] \,.
\end{eqnarray}
Since there is no Riemannian metric, the spectral invariants of  a non-Laplace type operator
are not expressed in terms of the invariants of the curvature. It is very important 
to develop the corresponding differential-geometric language 
based on the non-scalar leading symbol of a non-Laplace type operator.

Boundary value problems for non-Laplace type operators are even
more complicated.
For the Dirichlet boundary value problem the asymptotics are 
computed as follows (for details see \cite{avramidi06}).
Let $x=(r,\hat x)$ be the local coordinates near the boundary 
where $r$ is the normal geodesic distance to the boundary and
$\hat x^i$ are local coordinates on the boundary.
We decompose a covector $\xi\in T^*M$ as $\xi=(\omega,\hat \xi)$
where $\omega$ is a real number and $\hat \xi\in T^*\partial M$
is a covector on the boundary. 
Let  $\Phi(\lambda;\hat x,\hat\xi)$ be a function defined by
\be
\Phi(\lambda;\hat x,\hat \xi)=\int\limits_\RR
\frac{d\omega}{2\pi}\left\{H(0,\hat x,\omega,\hat\xi)-\lambda I\right\}^{-1}
\ee
and 
$\Psi(x,\xi)$ be a function defined by 
\be
\Psi(x,\xi)=\int\limits_{c-i\infty}^{c+i\infty}\frac{d\lambda}{2\pi i}
e^{-\lambda}\frac{\partial}{\partial \lambda}
\log\det \Phi(\lambda;\hat x,\hat\xi),
\ee
where $c$ is a sufficiently large positive real constant.
Then the heat trace boundary coefficient $A_1$ for the Dirichlet boundary conditions
has the form
\cite{avramidi06}
\be
A_1=-\sqrt{\pi}\int\limits_{\partial M}d\hat x\int\limits_{\RR^{n-1}}
\frac{d\hat \xi}{\pi^{(n-1)/2}}\Psi(\hat x,\hat\xi).
\ee

 
\section{Non-perturbative Spectral Asymptotics}
\setcounter{equation}0 

Let $M$ be a compact Riemannian manifold without boundary and
$\cV$ be a complex vector bundle over $M$ realizing a representation of the group
$G\times U(1)$. Let $\varphi$ be a section of the bundle $\cV$
and $\nabla$ be the total connection on the
bundle $S$ (including the $G$-connection as well as the $U(1)$-connection). Then the
commutator of covariant derivatives defines the curvatures
\be
[\nabla_\mu,\nabla_\nu]\varphi = (\cR_{\mu\nu} + iF_{\mu\nu})\varphi,
\ee
where $\cR_{\mu\nu}$ is the curvature of the $G$-connection and 
$F_{\mu\nu}$ is the curvature of the
$U(1)$-connection.

We assume that the $U(1)$-connection is parallel, that is,
\be
\nabla_\mu F_{\alpha\beta}=0.
\ee
This equation puts severe algebraic restriction on the curvature tensor
\be
R^\lambda{}_{\alpha\mu\nu}F_{\lambda \beta}
-R^\lambda{}_{\beta\mu\nu}F_{\lambda \alpha}
=0
\ee
and, therefore, gives powerful restriction on the holonomy group of the manifold.
For example, $F$ could be a simplectic form of a K\"ahler manifold.

Let $U(t;x,x')$ be the heat kernel of the Laplacian
$\Delta=g^{\mu\nu}\nabla_\mu\nabla_\nu$.
We rescale the curvature by
\be
F\mapsto \frac{1}{\varepsilon^2}F,
\ee
where $\varepsilon>0$ is a small positive parameter;
let $\Delta_\varepsilon$ be the rescaled Laplacian and 
$U_\varepsilon(\varepsilon t; x,x')$ be the rescaled heat kernel.
Then as $\varepsilon\to 0$ there is the 
{\it new asymptotic expansion} of the off-diagonal heat kernel
\cite{avramidi09d}
\be
U_\varepsilon(\varepsilon^2 t;x,x')\sim U_0(t;x,x')\sum_{k=0}^\infty \varepsilon^{k-n} t^{k/2}b_k(t;x,x')\,,
\ee
where 
\be
U_0(t;x,x')=(4\pi t)^{-n/2}\Delta^{1/2}(x,x')
\det\left(\frac{tiF}{\sinh(tiF)}\right)^{1/2}
\exp\left(-\frac{1}{4t}\left<u, tiF\coth(tiF)u\right>\right),
\ee
$F$ is the matrix
$F=(F_{\mu\nu})$
and $u^\mu$ are normal coordinates with origin at $x'$. 
Here the coefficients $b_k(t;x,x')$
are analytic functions of $t$ that depend on $F$ only in the dimensionless combination 
$tF$. Of course, for $t = 0$ they are equal to the standard heat kernel
coefficients, that is,
\be
b_{k}(0;x, x)=a_{k}(x,x'),
\ee
and, therefore, the odd-order coefficients at $t=0$ vanish,
$
b_{2k+1}(0;x,x') = 0. 
$
Moreover, the odd-order coefficients vanish also for any $t$ 
on the diagonal $x = x'$ \cite{avramidi09d}, that is,
\be
b^{\rm diag}_{2k+1}(t)=0
\ee
Then the asymptotic
expansion of the heat kernel diagonal and the heat trace are
\be
\Tr \exp(\varepsilon^2 t\Delta_\varepsilon)\sim (4\pi t)^{-n/2}
\sum_{k=0}^\infty \varepsilon^{k-n}t^{k/2} B_k(t)\,,
\ee
where
\be
B_k(t)=\int\limits_M d\vol \det\left(\frac{tiF}{\sinh(tiF)}\right)^{1/2}
\tr\;b^{\rm diag }_k(t)\,.
\ee
The coefficients $B_k(t)$ are {\it new spectral invariants} of the Laplacian.
They are differential polynomials
in the Riemann curvature tensor (and the curvature of the
$G$-connection) and its derivatives with universal coefficients depending in a
non-polynomial but analytic way on the curvature $F$, more precisely, on $tF$.

We explicitly computed the coefficients $b_k(t)$ (both off diagonal and the 
diagonal values) for $k=0,1,2,3,4$.
These functions generate all terms quadratic and linear in the Riemann
curvature and of arbitrary order in $F$ in the standard
heat kernel coefficients $a_k^{\rm diag}$.
In that sense, we effectively sum up the usual short time heat kernel
asymptotic expansion to all orders of the curvature $F$. 
The first two non-zero coefficients have the form
\cite{avramidi09d}
\bea
b_0^{\rm diag}(t) &=& I,
\\
b_2^{\rm diag}(t) &=& J_{\alpha\beta\mu\nu}(t)R^{\alpha\beta\mu\nu}I
+\frac{1}{2}H^{\mu\nu}(t)\cR_{\mu\nu},
\eea
where
\be
H(t)= \coth(tiF) -\frac{1}{itF}
\ee
and $J_{\alpha\beta\mu\nu}(t)$ is a more complicated tensor constructed from the matrix
$F$ that is analytic in $t$ 
(for more details see \cite{avramidi09d})
\be
J^{\alpha\beta}{}_{\mu\nu}(t)=\frac{1}{6}\delta^{\alpha}{}_{[\mu}\delta^{\beta}{}_{\nu]}
+O(t^2)
\ee


\section{Heat Determinant}
\setcounter{equation}0 

The existence of non-isometric isospectral manifolds
demonstrates that the spectrum alone does not determine the geometry. 
That is why it is worth studying new invariants 
that depend not only on the eigenvalues but also on the eigenfunctions, and,
therefore, contain much more information about the geometry of the manifold.

Let 
\be
L=-\Delta+Q
\ee
be a Laplace type operator acting on sections of 
a $N$-dimensional vector bundle $\cV$ over a $n$-dimensional 
compact Riemannian manifold $M$ without boundary. 
We define a new invariant (called the heat determinant)
by
(the detailed context and motivation for introducing this invariant is outlined
in \cite{avramidi14a})
\be
K(t)=\int\limits_{M\times M}dx\,dx'\; 
\det\left\{ \tr \left(U^*(t;x,x')\nabla_\mu\nabla_{\nu'}U(t;x,x')\right)\right\}\,,
\ee
where $U(t;x,x')$ is the heat kernel of the operator $L$.
This invariant has an interesting spectral representation.
Let $\{\lambda_k,\varphi_k\}_{k=1}^\infty$ be the eigenvalues and the eigensections
of the operator $L$ and
$\Phi^k_{l}$ be the one-forms defined by
\be
\Phi^k_{l}=\left<\varphi_k, D\varphi_l\right>
=\left<\varphi_k, \nabla_\mu\varphi_l\right> dx^\mu\,,
\ee
where $\left<\cdot,\cdot\right>$ is the fiber inner product,
and $D=d+\cA$ is the covariant exterior derivative.
Then the coefficients
\bea
\Psi^{k_1\dots k_n}_{l_1\dots l_n} 
&=&
\int\limits_M \Phi^{k_1}_{l_1}\wedge\cdots\wedge \Phi^{k_n}_{l_n}
\,
\eea
measure the correlations between the eigensections.
Then the invariant $K(t)$ takes the form
\cite{avramidi14a}
\be
K(t)=\frac{1}{n!}
\sum_{k_1,l_1,\dots , k_n, l_n=1}^\infty
\exp\left\{-t(\lambda_{k_1}+\lambda_{l_1}+\cdots+\lambda_{k_n}+\lambda_{l_n})\right\}
\left|\Psi_{l_1\dots l_n}^{k_1\dots k_n}\right|^2\,.
\ee

We prove in \cite{avramidi14a}
that there is an asymptotic expansion as $t\to 0$
\be
K(t)\sim 
\frac{1}{2}N^n(4\pi)^{-n^2}\left(\frac{\pi}{2n}\right)^{n/2}t^{-n\left(n+\frac{1}{2}\right)}
\sum_{k=0}^\infty t^{k/2} C_k,
\ee
where 
\be
C_k=\int\limits_M\; d\vol c_k,
\ee
and $c_k$ are differential polynomials in the Riemann
curvature, the curvature of the bundle connection and the potential $Q$
with some universal numerical coefficients that depend 
only on the dimensions $n$ and $N$. On manifolds without boundary all odd-order
coefficients vanish
\cite{avramidi14a}
\be
C_{2k+1}=0.
\ee
In particular,  
\be
C_0=\vol(M)
\ee
and
the coefficients  $c_2$ and $c_4$ are
computed explicitly in our paper \cite{avramidi14a}.


\section{Quantum Heat Traces}
\setcounter{equation}0

We initiate the study of new invariants of second-order
elliptic partial
differential operators acting on sections of vector bundles over 
compact Riemannian
manifolds without boundary. 
We draw a deep analogy
between the spectral invariants of elliptic operators and 
the (classical and quantum) statistical physics
\cite{landau69}.

In statistical physics one considers an ensemble of 
identical particles in thermodynamic equilibrium
at the temperature $T$ and the chemical potential $\mu$.
The average number of classical particles $n_k$ in the state with the energy
$E_k$ is determined by the
Boltzman distribution
\be
n_k=\exp[-\beta (E_k-\mu)],
\ee
where $\beta=1/T$ is the inverse temperature.
The quantum particles are indistinguishable \cite{landau69};
as the result,
there are two types of quantum particles, bosons and
fermions. The average number of bosons  
with energy $E_k$ is given by the
Bose-Einstein distribution
\be
n_{(b),k}=\frac{1}{\exp[\beta (E_k-\mu)]-1},
\ee
while the average number of fermions  
with energy $E_k$ is given by the
Fermi-Dirac distribution
\be
n_{(f),k}=\frac{1}{\exp[\beta (E_k-\mu)]+1}.
\ee
The energy spectrum $E_k$ is described by the Hamiltonian $H$.
The total number of particles determines the partition function
\be
Z(\beta,\mu)
=\Tr\exp[-\beta (H-\mu)],
\ee
in the classical case, 
and 
\be
Z_{b,f}(\beta,\mu)=\Tr\frac{1}{\exp[\beta (H-\mu)]+1}.
\ee
in the quantum case.
Recall that the Hamiltonian of a free particle of mass $m$ and momentum $p$ is
\be
H=\frac{p^2}{2m}
\ee
in the nonrelativistic case
and
\be
H=(p^2+m^2)^{1/2}
\ee
in the relativistic case (in the units with the speed of light $c=1$).

This motivates the study of generalized heat traces defined as follows
\cite{avramidi17}. 
Let $L$ be an elliptic self-adjoint positive partial differential operator
of second order acting on smooth section of a vector bundle $\cV$ over a closed manifold
$M$. Its square root,
\be
H=L^{1/2},
\ee
is an elliptic self-adjoint positive
{\it pseudo}-differential operator of first order
that plays the role of the Hamiltonian.
We interpret the classical heat trace 
\be
\Theta(\beta)=\Tr \exp(-\beta L)
\label{79fa}
\ee
as the partition function for the Boltzman distribution.
By analogy, we define 
the {\it relativistic heat 
trace}
\bea
\Theta_r(\beta) &=& \Tr \exp(-\beta H)
\eea
and the {\it quantum heat 
traces}
\bea
\Theta_b(\beta,\mu) &=&\Tr \frac{1}{\exp[\beta(H-\mu)]-1},
\\[5pt]
\Theta_f(\beta,\mu) &=&\Tr \frac{1}{\exp[\beta(H-\mu)]+1},
\eea
where $\mu$ is a parameter that plays the role of the chemical potential. 
We show that these new
invariants can be reduced to some integrals of the classical heat
trace
and compute the high-temperature asymptotics of these invariants as $\beta\to 0$.

We introduce a function $A_q$ of a complex variable $q$ defined by
the Mellin transform of the heat trace
\bea
A_q &=& (4\pi)^{n/2}\frac{1}{\Gamma(-q)}\int_0^\infty
dt\;t^{-q-1+n/2}\Theta(t).
\label{323fa}
\eea
Then we show (for details see \cite{avramidi91,avramidi17}) that for a positive operator $L$:
\begin{enumerate}
\item
the function $A_q$ is an entire function of $q$,
\item
its values
at non-negative integer points $A_k$, with $k\in\ZZ_+$,
are equal (up to a normalization factor)
to the standard heat trace
coefficients (\ref{240fa}), which are {\it locally computable}, 
\item
while the values
of the function $A_{k+1/2}$ at the half-integer points $k+1/2$, with $k$
a positive integer, as well
as the values of its derivative $A'_k=\partial_q A_q |_{q=k}$ at 
the positive integer points $k\in\ZZ_+$
are {\it new global invariants} that are {\it not locally computable}.
\end{enumerate}

We use the integral
\cite{prudnikov83}
\be
\exp(-x)=
(4\pi)^{-1/2}
\int_0^\infty
dt\;t^{-3/2}\exp\left(-\frac{1}{4t}-tx^2\right),
\label{31bb}
\ee
valid for $x\ge 0$,
to reduce the relativistic heat trace to the 
classical heat trace 
(\ref{79fa}), that is,
\be
\Theta_r(\beta)=
(4\pi)^{-1/2}
\int_0^\infty
dt\;t^{-3/2}\exp\left(-\frac{1}{4t}\right)\Theta(t\beta^2).
\label{32bb}
\ee
Next, by using eqs. (\ref{32bb}) and (\ref{323fa})  
we express
the relativistic heat trace in terms of the function $A_q$
via a Mellin-Barnes integral
\be
\Theta_r(\beta)=
2(4\pi)^{-(n+1)/2}
\frac{1}{2\pi i}
\int\limits_{c-i\infty}^{c+i\infty}dq\,\Gamma(-q)\Gamma\left[-q+(n+1)/2\right]
\left(\frac{\beta}{2}\right)^{2q-n}A_q.
\label{597a}
\ee

We compute the asymptotics of the relativistic heat trace $\Theta_r(\beta)$
as $\beta\to 0$ (for details of the calculation see \cite{avramidi17}).
We obtained
in even dimension $n=2m$,
\bea
\Theta_{r}(\beta)&\sim& 
\sum_{k=0}^\infty \beta^{2k-2m}b^{(1)}_k A_k
+\sum_{k=0}^\infty \beta^{2k+1}b^{(2)}_{k}A_{k+m+1/2},
\eea
and in odd dimension $n=2m+1$,
\bea
\Theta_{r}(\beta)&\sim& 
\sum_{k=0}^\infty\beta^{2k-2m-1}
b^{(3)}_kA_k
+\log\beta\sum_{k=0}^\infty 
\beta^{2k+1}b^{(4)}_{k}A_{k+m+1}
+\sum_{k=0}^\infty \beta^{2k+1}
b^{(5)}_{k}A'_{k+m+1},
\nonumber\\
\eea
and computed all numerical coefficients $b^{(i)}_k$.
Notice that the coefficients of the singular part containing the inverse powers of $\beta$
and the logarithm are locally computable.

Similarly, we express the quantum heat traces in terms of the
classical one
\cite{avramidi17}
\bea
\Theta_{b,f}(\beta,\mu) &=& \int\limits_0^\infty dt\; 
h_{b,f}\left(t,\beta\mu\right)
\Theta\left(t\beta^2\right),
\label{314zz}
\eea
where
\bea
h_f(t,\beta\mu) &=& (4\pi)^{-1/2}t^{-3/2}
\sum_{k=1}^\infty (-1)^{k+1} k\exp\left(-\frac{k^2}{4t}+k\beta\mu\right),
\\
h_b(t,\beta\mu) &=& (4\pi)^{-1/2}t^{-3/2}
\sum_{k=1}^\infty k\exp\left(-\frac{k^2}{4t}+k\beta\mu\right),
\eea
which reduces the calculation of the quantum heat traces to the calculation of the
classical one (\ref{79fa}).
This gives the Mellin-Barnes representation of the quantum heat traces
\cite{avramidi17}
\bea
\Theta_{b,f}(\beta,\mu) &=& 
2(4\pi)^{-(n+1)/2}
\frac{1}{2\pi i}\int\limits_{c-i\infty}^{c+i\infty}dq\;
\Gamma(-q)\Gamma\left[-q+(n+1)/2\right]\left(\frac{\beta}{2}\right)^{2q-n}
\nonumber\\
&&\times
F_{b,f}(n-2q,\beta\mu)A_q,
\label{597abb}
\eea
where $c<0$ and
\bea
F_b(s,{\beta\mu})&=&\sum_{k=1}^\infty\frac{e^{k{\beta\mu}}}{k^{s}}
=\frac{1}{\Gamma(s)}\int_0^\infty dt\;\frac{t^{s-1}}{e^{t-{\beta\mu}}-1},
\\
F_f(s,{\beta\mu})&=&\sum_{k=1}^\infty(-1)^{k+1}\frac{e^{k{\beta\mu}}}{k^{s}}
=\frac{1}{\Gamma(s)}\int_0^\infty dt\;\frac{t^{s-1}}{e^{t-{\beta\mu}}+1}.
\eea

We compute their asymptotics as $\beta\to 0$
\cite{avramidi17}. 
For 
$\mu=0$ we obtain an asymptotic expansion as $\beta\to 0$:
in even dimension $n=2m$,
\bea
\Theta_{f}(\beta,0)&\sim& 
\sum_{k=0}^m \beta^{2k-2m}c^{(1)}_k A_k
+\sum_{k=0}^\infty \beta^{2k+1}c^{(2)}_{k}A_{k+m+1/2},
\\
\Theta_{b}(\beta,0) &=&
\sum_{k=0}^m\beta^{2k-2m}c^{(3)}_kA_k
+\sum_{k=-1}^\infty \beta^{2k+1}c^{(4)}_{k}A_{k+m+1/2},
\eea
and in odd dimension $n=2m+1$,
\bea
\Theta_{f}(\beta,0)&\sim& 
\sum_{k=0}^\infty\beta^{2k-2m-1}
c^{(5)}_kA_k
+\log\beta\sum_{k=0}^\infty 
\beta^{2k+1}c^{(6)}_{k}A_{k+m+1}
+\sum_{k=0}^\infty \beta^{2k+1}
c^{(7)}_{k}A'_{k+m+1},
\nonumber\\
\\
\Theta_{b}(\beta,0) &=&
\sum_{k=0}^{m-1} 
\beta^{2k-2m-1}
c^{(8)}_kA_k
+\log\beta\sum_{k= -1}^\infty 
\beta^{2k+1}
c^{(9)}_{k}A_{k+m+1}
+\sum_{k=-1}^\infty 
\beta^{2k+1}c^{(10)}_{k}A'_{k+m+1}.
\nonumber\\
\eea
and computed all numerical coefficients $c^{(i)}_k$
in \cite{avramidi17}.


\section{Relative Spectral Invariants}
\setcounter{equation}0 


As we already mentioned above the spectrum of a single elliptic partial
differential operator does not
determine the geometry of a manifold. 
One could ask the natural question whether the spectral information of two 
elliptic differential operators determines the geometry. 
Our goal is to define new relative spectral invariants that depend both on the eigenvalues
and the eigensections of both operators and contain much more information about geometry. 
Such relative spectral invariants appear naturally, in particular, in the study of particle
creation in quantum field theory and quantum gravity 
\cite{dewitt75,birrell80,avramidi20a,avramidi20b}.

Let $L_\pm$ be two self-adjoint elliptic second-order partial 
differential operators acting on smooth sections of the vector bundle $\cV$
over a closed manifold $M$
with a positive definite scalar leading symbols of Laplace type,
\be
L_\pm = -g_\pm^{-1/4}(\partial_i+\cA^\pm_i)g_\pm^{1/2}g_\pm^{ij}
(\partial_j+\cA^\pm_j)g_\pm^{-1/4}+Q_\pm.
\ee
Here $g_{ij}^\pm$ are two metrics, $\cA^\pm_i$ are two connection
one-forms and $Q_\pm$ are two endomorphisms;
also $g^{ij}_\pm$ are the inverse metrics and $g_\pm=\det g^\pm_{ij}$.

We assume that $\cV$ is a Clifford bundle.
Let $D_\pm$ be two self-adjoint elliptic first-order partial differential 
operators acting on smooth sections of the vector bundle $\cV$ of
Dirac type, 
\be
D_\pm = g_\pm^{1/4}i\gamma^a e^j_{\pm a}(\partial_j+\cA_j^\pm)g_\pm^{-1/4}+S_\pm,
\ee
where $\gamma^a$ are the Dirac matrices satisfying
\be
\gamma_a\gamma_b+\gamma_b\gamma_a=2\delta_{ab}I,
\ee
$e^i_{\pm a}$ are two orthonormal frames for the metrics $g_\pm^{ij}$
satisfying
\be
g^{ij}_\pm = \delta^{ab}e^i_{\pm a}e^i_{\pm b},
\ee
$\cA_j^\pm$ are two connections and $S_\pm$ are two endomorphisms.
We require that the endomorphisms $S_\pm$ commute with the Dirac matrices
so that the square of the Dirac type operators,  $D_\pm^2$, are operators
of Laplace type.

The spectral information about the operators $L_\pm$ and $D_\pm$
is contained in the classical heat traces
\bea
\Theta_\pm(t) &=& \Tr \exp(-tL_\pm),
\label{237ssb}
\\
H_\pm(t) &=& \Tr D_\pm \exp(-tD_\pm^2).
\label{12via}
\eea
The relative spectral information is contained in the
 {\it relative spectral invariants}
\cite{avramidi20b}
\bea
\Psi(t,s) &=&
\Tr\left\{\exp(-tL_+)-\exp(-tL_-)\right\}
\left\{\exp(-sL_+)-\exp(-sL_-)\right\},
\label{220ssb}
\\[10pt]
\Phi(t,s) &=&
\Tr\left\{D_+\exp(-tD^2_+)-D_-\exp(-tD^2_-)\right\}
\left\{D_+\exp(-sD^2_+)-D_-\exp(-sD^2_-)\right\},
\nonumber\\
\label{17via}
\eea
which can be expressed in terms of the
{\it combined heat traces}
\bea
X(t,s) &=& \Tr\left\{\exp(-tL_+)\exp(-s L_-)\right\}
\nonumber\\
&=&
\int\limits_{M\times M}dx\;dx'\;\tr\left\{ U_+(t;x,x')U_-(s;x',x)\right\},
\label{238ccx}
\\
Y(t,s) &=& \Tr\left\{ D_+\exp(-tD^2_+)D_-\exp(-sD^2_-)\right\}
\nonumber\\
&=& 
\int\limits_{M\times M}dx\;dx'\;\tr\left\{
 D_+U_+(t;x,x')D_-U_-(s;x',x)\right\}.
\label{18ssb}
\eea

We study the asymptotics of the 
combined heat traces (\ref{238ccx}) and (\ref{18ssb}).
We define the time-dependent
metric $g_{ij}=g_{ij}(t,s)$ as the inverse of the matrix
\be
g^{ij}=tg_+^{ij}+sg_-^{ij},
\label{113via}
\ee
with $t,s> 0$; throughout the paper we use the notation
$g=\det g_{ij}$ for the determinant of the metric. 
Also, we define the time-dependent connection 
$\cA_i=\cA_i(t,s)$
by
\be
\cA_i=g_{ij}\left(tg^{jk}_+\cA^+_k+sg^{jk}_-\cA^-_k\right).
\label{114via}
\ee
and the vectors
\be
\cC^\pm_i = \cA^\pm_i-\cA_i.
\label{115via}
\ee

\begin{theorem}
\label{theorem1}
There are asymptotic expansions as $\varepsilon\to 0$
\bea
X(\varepsilon{}t,\varepsilon{}s) &\sim& (4\pi\varepsilon)^{-n/2}
\sum_{k=0}^\infty \varepsilon^{k} B_k(t,s),
\label{1zaab}
\\
Y(\varepsilon{}t,\varepsilon{}s) &\sim&
(4\pi\varepsilon)^{-n/2}
\sum_{k=0}^\infty \varepsilon^{k-1} 
C_k(t,s),
\label{15zaac}
\eea
where 
\bea
B_k(t,s) &=& \int\limits_M dx\; g^{1/2}(t,s)b_k(t,s),
\\
C_k(t,s) &=& \int\limits_M dx\; g^{1/2}(t,s)c_k(t,s).
\eea
\begin{enumerate}
\item 
The coefficients $b_k(t,s)$ and $c_k(t,s)$
are scalar invariants built polynomially from the covariant
derivatives (defined with respect to the metric $g_{ij}$ and the
connection $\cA_i$)
of the metrics $g^\pm_{ij}$, the vectors $\cC^\pm_i$ and
the potentials $Q_\pm$ and $S_\pm$.

\item
The coefficients $b_k(t,s)$
are homogeneous functions of $t$ and $s$
of degree $k$ and the coefficients $c_k(t,s)$ are 
homogeneous
functions of $t$ and $s$
of degree $(k-1)$.
\end{enumerate}
\end{theorem}

{\it Proof:} This theorem is proved in \cite{avramidi20b}.


In particular, the first coefficients of the asymptotic expansion of the combined
heat traces are
\cite{avramidi20b}
\bea
b_0(t,s) &=& N,
\label{132via}
\\
c_0(t,s) &=&
N\frac{1}{2}\delta^{ab}
e_{+a}^{i}g_{ij}(t,s)e_{-b}^{j},
\label{138via}
\eea
where $N=\dim \cV$.


\section{Bogolyubov Invariant}
\setcounter{equation}0 


To motivate the definition of the Bogolyubov invariant 
we consider Klein-Gordon and Dirac fields on a globally
hyperbolic manifold $\cM$ with compact Cauchy surfaces $M_t$, $t\in\RR$, satisfying an
assumption on good asymptotic behaviour as 
$t\to \pm\infty$. The time-dependence of the operators leads in quantum field theory
to the phenomenon of particle creation. We introduce below an invariant,
called a Bogolyubov invariant,
that, in a relevant asymptotic regime, describes the renormalized number of 
created particles. This invariant is closely related to the relative
spectral invariants discussed above. A much more detailed discussion is
presented in \cite{avramidi20a}.

Let $\cV$ be a twisted spin-tensor bundle.
Let $\xi$ and $\eta$ be two self-adjoint anti-commuting 
involutive endomorphisms of the bundle ${\cal V}$,
\be
\xi^2=\eta^2=I,
\qquad
\xi^*=\xi, 
\qquad
\eta^*=\eta,
\qquad
\xi\eta=-\eta\xi.
\ee
Let $D_\pm$ be a {self-adjoint} 
elliptic {first-order} partial differential operators
of Dirac type
acting on smooth sections of the bundle $\cV$ 
that anti-commute with $\eta$ and commute with $\xi$,
\be
D_\pm\eta=-\eta D_\pm,
\qquad
D_\pm\xi=\xi D_\pm.
\ee
Suppose that the square of the operators $D_\pm$ 
are Laplace type operators
\be
D^2_\pm=H_\pm.
\ee
Then the operators
$(D_\pm+m\eta)$ with a mass parameter $m$
are Dirac type operators
whose square are Laplace type operators
\be
(D_\pm+m\eta)^2=H_\pm+m^2I.
\label{214xxx}
\ee

The bosonic and fermionic Bogolyubov invariants
are defined by the following traces
\cite{avramidi20a}
\bea
B_b(\beta)&=&\Tr
\left\{ \frac{1}{\exp(\beta\omega_+)+1}
-\frac{1}{\exp(\beta\omega_-)+1}
\right\}
\left\{ \frac{1}{\exp(\beta\omega_+)-1}
-\frac{1}{\exp(\beta\omega_-)-1}
\right\},
\nonumber\\
\label{311xxa}
\\
B_f(\beta) &=& \beta^2\Tr
\left\{
\frac{(D_++m\eta)}{\sinh (\beta\omega_+)}
-\frac{(D_-+m\eta)}{\sinh (\beta\omega_-)}
\right\}^2.
\label{318xxa}
\eea
where
\be
\omega_\pm=(H_\pm+m^2)^{1/2},
\ee

The Bogolyubov invariants can be expressed in terms of the
the {\it relative spectral invariants}
$\Psi(t,s)$ and $\Phi(t,s)$ defined in
(\ref{220ssb}) and (\ref{17via}).
The heat trace representation for the Bogolyubov invariant
can be obtained as follows \cite{avramidi20a}.
Let $h_{b,f,0}$ be the functions defined by
\bea
h_f(t) &=& 
(4\pi)^{-1/2}t^{-3/2}
\sum_{k=1}^\infty (-1)^{k+1} k\exp\left(-\frac{k^2}{4t}\right),
\\
h_b(t) &=&
(4\pi)^{-1/2}t^{-3/2}
\sum_{k=1}^\infty k\exp\left(-\frac{k^2}{4t}\right),
\\
h_0(t) &=& 
(4\pi)^{-1/2}t^{-3/2}
\sum_{k=0}^\infty \left( 2k+1\right)
\exp\left(-\frac{\left(2k+1\right)^2}{4t}\right).
\eea
Then, for the bosonic case
we get from (\ref{311xxa})
\cite{avramidi20a}
\be
B_b(\beta) = 
\int\limits_0^\infty dt\int\limits_0^\infty ds\;
h_f\left(t\right)
h_b\left(s\right)
\exp\left(-m^2\beta^2(s+t)\right)\Psi\left(\beta^2t,\beta^2s\right),
\label{413xxa}
\ee
where  
\be
\Psi(t,s)=
\Tr\left[\exp(-tH_+)-\exp(-tH_-)\right]
\left[\exp(-sH_+)-\exp(-sH_-)\right].
\label{248xnxa}
\ee
For the fermionic case we get from (\ref{318xxa})  \cite{avramidi20a}
\bea
B_f(\beta) &=&
\int\limits_0^\infty dt\int\limits_0^\infty ds\;\,
h_0\left(s\right)
h_0\left(t\right)
\exp\left(-m^2\beta^2(s+t)\right)
\nonumber\\
&&\times
\beta^2\left\{\Phi\left(\beta^2t,\beta^2s\right)
+m^2\Psi\left(\beta^2t,\beta^2s\right)\right\},
\label{414xxa}
\eea
where 
\be
\Phi(t,s)=
\Tr\left[D_+\exp(-tH_+)-D_-\exp(-tH_-)\right]
\left[D_+\exp(-sH_+)-D_-\exp(-sH_-)\right].
\label{248xxa}
\ee
It is easy to see that  the integrals
for the Bogolyubov invariant converge both as $t,s\to 0$
and (for sufficiently large $m^2$) also as $t,s\to \infty$.


Now we can compute the asymptotics of Bogolyubov invariant as $\beta\to 0$.
(We list below the results only; for the details of the calculation see
\cite{avramidi20a}.)
We obtain om odd dimension
$n=2m+1$
\be
B_b(\beta)\sim
\sum_{k=0}^\infty \beta^{2k-n}c_k^{(1)}
+\sum_{k=0}^\infty \beta^{2k}c_k^{(2)},
\label{712saaz}
\ee
and in the even dimension $n=2m$
\be
B_b(\beta)\sim
\sum_{k=0}^{m-1} \beta^{2k-n}c_k^{(3)}
+\sum_{k=0}^{\infty} \beta^{2k}c_k^{(4)}
+\log\beta^2\;\sum_{k=0}^\infty \beta^{2k}c_k^{(5)}.
\label{715saz}
\ee
Here $c_k^{(i)}$ are some coefficients.
Notice that the coefficients $c^{(1)}_k$ of the all
odd powers of $\beta$ as well as
the coefficients $c^{(3)}_k$ and $c^{(5)}_k$ 
of the singular part and the logarithmic part
are locally computable invariants whereas the coefficients
$c^{(2)}_k$ of the even non-negative powers of $\beta$ 
as well as the coefficients $c^{(4)}_k$ of the regular part
are non-locally computable global invariants.
The leading asymptotics have the form
\be
B_b(\beta)=\beta^{-n}c_0^{(1)}+\beta^{-n+2}c^{(1)}_1 +O(\beta^{-n+4})+O(\log\beta).
\ee

Similarly, for the Dirac operator we obtain:
in the odd dimension $n=2m+1$:
\be
B_f(\beta)\sim
\sum_{k=0}^\infty \beta^{2k-n}d_k^{(1)}
+\sum_{k=0}^\infty \beta^{2k}d_k^{(2)},
\label{712saax}
\ee
and in the even dimension $n=2m$
\be
B_f(\beta)\sim
\sum_{k=0}^{m-1} \beta^{2k-n}d_k^{(3)}
+\sum_{k=0}^{\infty} \beta^{2k}d_k^{(4)}
+\log\beta^2\;\sum_{k=0}^\infty \beta^{2k}d_k^{(5)},
\label{715sazd}
\ee
Here the coefficients $d^{(1)}_k$ of the all
odd powers of $\beta$
as well as the coefficients 
 $d^{(3)}_k$ and $d^{(5)}_k$ 
of the singular part and the logarithmic part
are locally computable invariants whereas the coefficients
$d^{(2)}_k$ of the even non-negative powers of $\beta$ 
as well as the coefficients $d^{(4)}_k$ of the regular part are
non-locally 
computable global invariants.
The leading asymptotics have the form
\be
B_f(\beta)=\beta^{-n}d_0^{(1)}+\beta^{-n+2}d^{(1)}_1 +O(\beta^{-n+4})+O(\log\beta).
\ee


\section{Heat Semigroups on Weyl Algebra}
\setcounter{equation}0 

Below we present a particular model when the combined heat traces
can be computed exactly based only on the algebraic properties
of the operators involved \cite{avramidi21}.

Let $x^i$ be the coordinates of the Euclidean space $\RR^n$
and $\partial_i$ be the corresponding partial derivatives.
More precisely, we consider the partial derivative
operators acting on the space $C^\infty_0(\RR^n)$
of { smooth functions of compact support}. Recall that
this space is dense in the Hilbert space $L^2(\RR^n)$,
which defines the extension of these operators to the 
whole Hilbert space $L^2(\RR^n)$. 
Moreover, the partial derivatives are
unbounded (essentially)
{ anti-self-adjoint} operators in this Hilbert space.
Then the Weyl algebra (the universal enveloping algebra of the Heisenberg algebra)
is simply
the ring of all differential operators with polynomial coefficients.

We define the anti-self-adjoint operators $\nabla_k$
of the form 
\be
\nabla_k=\partial_k-\frac{1}{2}i\cR_{kj}x^j.
\label{67igax}
\ee
Given a positive matrix $g$ we introduce an operator
called the { Laplacian} by 
\be
\Delta_g=g^{ij}\nabla_i\nabla_j,
\label{delta}
\ee
where $g^{ij}$ is the inverse matrix.

We consider two Laplacians $\Delta_\pm$ defined with two
different matrices $\cR^\pm_{ij}$ and two different matrices
$g^\pm_{ij}$ and study the convolution of the heat semigroups
\\
$
\exp(t\Delta_+)\exp(s\Delta_-),
$
more precisely,
the integral kernel of the
product of the corresponding heat semigroups
\bea
U(t,s;x,x') &=&\exp(t\Delta_+)\exp(s\Delta_-)\delta(x-x').
\eea


The $(2n+1)$ operators
$(\nabla_1,\dots,\nabla_n, x^1,\dots,x^n,i)$ 
form the Lie algebra $\hfrak_n$ of the Heisenberg group $H_{2n+1}$
with the commutation relations
\bea
[\nabla_k,x^j]&=&\delta^j_k,
\\{}
[\nabla_k,\nabla_j] &=& i\cR_{kj}\,,
\label{45igazx}
\\{}
[\nabla_k,i]&=&[x^k,i]=0.
\label{nabla-comm}
\eea
The universal enveloping algebra $U(\hfrak_n)$ of the Heisenberg algebra 
(called the Weyl algebra)
is the set of all polynomials in these operators subject to these
commutation relations.

The operator $\exp\left<\xi,\nabla\right>: C^\infty_0(\RR^n)\to C^\infty_0(\RR^n)$
acts by
\bea
\left(\exp\left<\xi,\nabla\right>f\right)(x)
&=&
\exp\left(-\frac{1}{2}\left<\xi,i\cR x\right>\right)
f(x+\xi)\,.
\label{612igaxb}
\eea
and 
is an isometry \cite{avramidi21}.

First, we show that the heat semigroup can be represented
as a non-commutative Gaussian integral
\cite{avramidi21}
\be
\exp(t\Delta_g) = (4\pi)^{-n/2}\Omega(t) 
\int\limits_{\RR^n} d\xi
\exp\left\{-\frac{1}{4}\left<\xi,D(t)\xi\right>
\right\}
\exp\left<\xi,\nabla\right>\,,
\label{858viax}
\ee
where $D(t)=(D_{ij})$
is a symmetric matrix defined by
\be
D(t) 
=i\cR\coth\left(tg^{-1}i\cR\right)\,,
\ee
and 
\bea
\Omega(t) &=& 
\det \left(g^{-1}\frac{\sinh(tg^{-1}i\cR)}{g^{-1}i\cR}\right)^{-1/2}.
\eea

Next, we consider two sets of such
operators $\nabla_i^+$ and $\nabla_j^-$
forming
the Lie algebra
\bea
[\nabla^+_i,\nabla^+_j] &=& i\cR^+_{ij},
\\{}
[\nabla^-_i,\nabla^-_j] &=& i\cR^-_{ij},
\\{}
[\nabla^+_i,\nabla^-_j] &=& i\cR_{ij}.
\label{843viazy}
\eea
where
\be
\cR_{ij}=\frac{1}{2}\left(\cR^+_{ij}+\cR^-_{ij}\right).
\label{84igax}
\ee

Then we compute the product of the semigroups $\exp(t\Delta_+)\exp(s\Delta_-)$.
We define the operators $\nabla_i$ and $X_i$ by
\bea
\nabla_i&=&\frac{1}{2}(\nabla^+_i+\nabla^-_i),
\label{351via}
\\
X_i&=&\nabla^+_i-\nabla^-_i.
\label{352viam}
\eea
Then we prove a highly non-trivial representation
(for details of the proof see \cite{avramidi21})
\bea
&&\exp(t\Delta_+)\exp(s\Delta_-)
=
(4\pi)^{-n/2}
\Omega(t,s)
\exp\left<X,D^{-1}(t,s)X\right>
\label{490viaxc}
\\
&&\qquad\times
\int\limits_{\RR^{n}} d\xi\,\exp\left\{
-\frac{1}{4}\left<\xi, H(t,s)\xi\right>
-\frac{1}{2}\left<\xi,Z^T(t,s)D^{-1}(t,s)X\right>
\right\}
\exp\left<\xi, \nabla\right>,
\nonumber
\eea
where $T_\pm(t)$, $D(t,s)$, $Z(t,s)$ and $H(t,s)$ are the  
matrices defined by
\bea
T_\pm(t)&=&D_\pm(t)+i\cR,
\\
D(t,s)&=& D_+(t)+D_-(s),
\label{762igazx}
\\
Z(t,s) &=& D_+(t)-D_-(s)-2i\cR_-,
\label{491viacx}
\\
H(t,s) &=& \frac{1}{4}\left(D(t,s)-Z^T(t,s)D^{-1}(t,s)Z(t,s)\right),
\label{859igaxy}
\eea
and 
\be
\Omega(t,s) = 
\left(\frac{\det T_+(t)\det T_-(s)}{\det D(t,s)}\right)^{1/2}.
\label{2128iganx}
\ee


This enables us to compute the convolution of the heat kernels
\cite{avramidi21}
\bea
U(t,s;x,x') &=&
(4\pi)^{-n/2}\left(\det B(t,s)\right)^{1/2}
\\
&&
\times
\exp\left\{
-\frac{1}{4}\left<x, A_+(t,s)x\right>
-\frac{1}{4}\left<x', A_-(t,s) x'\right>
+\frac{1}{2}\left<x,B(t,s)x'\right>
\right\},
\label{820iganm}
\nonumber
\eea
where $A_\pm(t,s)$ and $B(t,s)$ be matrices defined by
\bea
A_+(t,s) &=& D_+(t)-T_+^T(t)D^{-1}(t,s)T_+(t),
\\
A_-(t,s) &=& D_-(s)-T_-^T(s)D^{-1}(t,s)T_-(s),
\\
B(t,s)&=& T_+(t)D^{-1}(t,s)T_-(s).
\eea




\end{document}